\documentclass[aip,notitlepage, floatfix, jcp, preprint]{revtex4-1}
\usepackage{amssymb}
\usepackage{amsfonts}
\usepackage{amsmath}
\usepackage{graphicx}
\usepackage{enumerate}
\usepackage[usenames,dvipsnames]{color}
\usepackage{ulem}

\input{tcilatex}
\begin{document}

\title{Nucleation of colloids and macromolecules in a finite volume}
\author{James F. Lutsko}
\affiliation{Center for Nonlinear Phenomena and Complex Systems, Code Postal 231,
Universit\'{e} Libre de Bruxelles, Blvd. du Triomphe, 1050 Brussels, Belgium}
\email{jlutsko@ulb.ac.be}
\homepage{http://www.lutsko.com}

\begin{abstract}
A recently formulated description of homogeneous nucleation for Brownian
particles in the over-damped limit based on fluctuating hydrodynamics is
used to determine the nucleation pathway, characterized as the most likely
path (MLP), for the nucleation of a dense-concentration droplet of globular
protein from a dilute solution in a small, finite container. The
calculations are performed by directly discretizing the equations for the
MLP and it is found that they confirm previous results obtained for infinite
systems: the process of homogeneous nucleation begins with a
long-wavelength, spatially-extended concentration fluctuation that it
condenses to form the pre-critical cluster. This is followed by a classical
growth processes. The calculations show that the post-critical growth
involves the formation of a depletion zone around the cluster whereas no
such depletion is observed in the pre-critical cluster. The approach
therefore captures dynamical effects not found in classical Density
Functional Theory studies while consistently describing the formation of the
pre-critical cluster.
\end{abstract}

\date{\today }
\maketitle

\section{Introduction}

Recently, the process of diffusion-limited nucleation has been formulated
within the context of fluctuating hydrodynamics\cite{Lutsko_JCP_2011_Com,
Lutsko_JCP_2012_1, Lutsko_JCP_2012_2}. This approach gives a unified
description of the formation and growth of an unstable, pre-critical nucleus
by means of thermal fluctuations as well as of the post-critical growth
driven by a free energy gradient. In this dynamical approach to the
description of nucleation, the fundamental quantity is a hydrodynamic field,
the local concentration\cite{Note1}. It can be used to derive the elements of Classical Nucleation Theory (CNT)%
\cite{Lutsko_JCP_2012_1, Lutsko_CNT} when the concentration-field is
parameterized using the usual capillary approximation that is implicit in
CNT. More complex parameterizations are possible giving thereby a more
fine-grained description of the nucleation pathway. By this means, it was
found that the dynamical theory predicts an unexpected, non-classical
pathway wherein nucleation is initiated by a long-wavelength,
small-amplitude concentration fluctuation. Spatial localization of the mass
in the fluctuation leads to the formation of a relatively large pre-critical
cluster. This is followed by a second stage consisting of growth of the
pre-critical cluster in accordance with more classical approaches (CNT, see,
e.g., Ref. \onlinecite{Kashchiev}, Density functional theory (DFT), see,
e.g., Refs.%
\onlinecite{Evans79, OxtobyEvans, talanquer:5190,
LutskoAdvChemPhysDFT}, ...) One advantage of the dynamical approach is that
it can be used to determine the relative likelihood of different proposed
nucleation pathways and it was found that the non-classical path is much
more likely than those of more classical approaches, such as DFT, which
inevitably show cluster formation as a spatially-localized processes\cite%
{Lutsko_JCP_2012_2}.

All of the results described above were obtained for an infinite system.
This leaves open the question as to whether the non-classical path will
dominate in a finite system more typical of a real experiment. Since the
physics of the non-classical path is rooted in mass conservation, it seems
possible that in a finite system with finite mass, a different result could
emerge. This could be particularly relevant in very small systems such as
those typical of nano-fluidics, biological processes and lab-on-a-chip
applications. A second question left open in previous studies is whether the
use of parameterized profiles, no matter how complex, might be prejudicing
the results. The purpose of this paper is to address both of these issues.
Calculations based on the dynamical description of nucleation are presented
for the nucleation of a bubble of dense protein-rich solution from a
mother-phase of dilute protein solution in a small, finite volume. The
calculations are performed by directly integrating a discretized version of
the field equations that determine the most likely path in
concentration-field space connecting the initial, nearly uniform dilute
phase and a final state consisting of a large, stable droplet of dense
solution within a small spherical cavity. It is found that the two-step
mechanism is preserved in modified form and that further details of the
process that were inaccessible with the parameterized models, such as the
formation of a depletion zone during post-critical growth, are evident.
Further, the additional resolution afforded by the direct discretization of
the equations governing the nucleation pathway, as opposed to the use of
parameterizations, allows to further distinguish the results of the
dynamical theory from those of more standard DFT-based approaches.

The next section presents the basic theory describing nucleation in terms of
a temporally evolving concentration field. The stochastic description is
discretized and the determination of the initial state, critical cluster,
final state and most likely pathway between them is formulated. The third
section presents calculations for the nucleation of protein-rich model
globular protein from a dilute mother phase. The paper concludes with a
discussion of the results compared to those obtained by other approaches and
with some remarks concerning open questions.

\section{Theory}

\subsection{Stochastic evolution of the concentration}

The fundamental quantity in the present formulation is the local number
concentration $\rho \left( \mathbf{r}\right) $. In the over-damped limit,
the evolution of the local concentration is governed by the stochastic
differential equation\cite{Lutsko_JCP_2011_Com,Lutsko_JCP_2012_1} (SDE)%
\begin{equation}  \label{D0}
\frac{\partial \rho \left( \mathbf{r};t\right) }{\partial t}=D\mathbf{\nabla 
}\rho \left( \mathbf{r};t\right) \mathbf{\nabla }\left( \frac{\delta F\left[
\rho \right] }{\delta \rho \left( \mathbf{r}\right) }\right) _{\rho \left( 
\mathbf{r}\right) \rightarrow \rho \left( \mathbf{r};t\right) }+\mathbf{%
\nabla }\sqrt{2D\rho \left( \mathbf{r};t\right) }\mathbf{\xi }\left( \mathbf{%
r};t\right)
\end{equation}%
where the noise, $\mathbf{\xi }\left( \mathbf{r};t\right) $, is
delta-correlated in both space and time. Even though this SDE involves a
state-dependent noise amplitude, it can be shown\cite{Lutsko_JCP_2012_1}
that it is Ito-Stratonovich equivalent so that there is no ambiguity in its
interpretation. The functional $F\left[ \rho \right] $ is a coarse-grained
free energy. In general, it is not equivalent to the free energy functional
of DFT as the latter is an equilibrium quantity. However, it is generally
assumed that the coarse-grained quantity is well-approximated by a
mean-field approximation. Just to make clear the connection, one expects
that averaging Eq.(\ref{D0}) over the noise would give something like the
usual DDFT form 
\begin{equation}
\frac{\partial \left\langle \rho \left( \mathbf{r};t\right) \right\rangle }{%
\partial t}=\widetilde{D}\mathbf{\nabla }\left\langle \rho \left( \mathbf{r}%
;t\right) \right\rangle \mathbf{\nabla }\left( \frac{\delta \widetilde{F}%
\left[ \rho \right] }{\delta \rho \left( \mathbf{r}\right) }\right) _{\rho
\left( \mathbf{r}\right) \rightarrow \left\langle \rho \left( \mathbf{r}%
;t\right) \right\rangle }
\end{equation}%
where, now, $\widetilde{F}\left[ \rho \right] $ is indeed the equilibrium
free energy functional of DFT. For spherically symmetric problems, the SDE
can be reduced to 
\begin{equation}  \label{D1}
\frac{\partial m\left( r;t\right) }{\partial t}=D4\pi r^{2}\rho \left(
r;t\right) \frac{\partial }{\partial r}\left( \frac{\delta F\left[ \rho %
\right] }{\delta \rho \left( \mathbf{r}\right) }\right) _{\rho \left( 
\mathbf{r}\right) \rightarrow \rho \left( r;t\right) }+\sqrt{8\pi r^{2}D\rho
\left( r;t\right) }\mathbf{\xi }\left( r;t\right)
\end{equation}%
where the spherically symmetric noise, $\mathbf{\xi }\left( r;t\right) $, is
delta-correlated in both variables\cite%
{Lutsko_JCP_2011_Com,Lutsko_JCP_2012_1}. Here, the quantity $m\left(
r;t\right) $ is the total mass within a ball of radius $r$,%
\begin{equation}
m\left( r;t\right) =\int_{0}^{r}4\pi r^{\prime 2}\rho \left( r^{\prime
};t\right) dr^{\prime }.
\end{equation}%
For the free energy functional, a simple squared-gradient model will be used%
\begin{equation}
F\left[ \rho \right] =\int \left\{ f\left( \rho \left( \mathbf{r}\right)
\right) +\frac{1}{2}K\left( \mathbf{\nabla }\rho \left( \mathbf{r}\right)
\right) ^{2}\right\} d\mathbf{r}
\end{equation}%
where $f\left( \rho \right) $ is the bulk Helmholtz free energy per unit
volume. Finally, it is interesting to note that the equation of motion can
be written in the particularly simple and illuminating form%
\begin{equation}
\frac{\partial m\left( r;t\right) }{\partial t}=-D\frac{\partial m\left(
r;t\right) }{\partial r}\left. \frac{\delta F\left[ \rho \right] }{\delta
m\left( r\right) }\right\vert _{m\left( r\right) \rightarrow m\left(
r;t\right) }+\sqrt{2D\frac{\partial m\left( r;t\right) }{\partial r}}\mathbf{%
\xi }\left( r;t\right)
\end{equation}
This shows that, in terms of the mass variable, the dynamics is
gradient-driven with a fluctuation-dissipation relation.

\subsection{Discretization}

\subsubsection{Variables}

In general, SDE's of the form given here are only well-defined once they are
discretized. Furthermore, discretization is necessary for practical
calculations. Here, space will be discretized via points $r_{i}=i\Delta$ and
the boundary of the finite volume of the container under consideration will
be at $r_{N}=\left( N+1/2\right) \Delta$. The masses at those points are $%
m\left( r_{i}\right) \equiv m_{i}$ for $0\leq i\leq N+1$ and, of course, $%
m_{0}=0$. The concentration will be represented at intermediate points as%
\begin{equation}
\rho\left( r_{i}+\frac{1}{2}\Delta\right) \equiv\rho_{i+1/2},\;0\leq i\leq N
\end{equation}
so that the relation with the mass is 
\begin{equation}
m_{i}=\Delta\sum_{j=0}^{i-1}4\pi
r_{j+1/2}^{2}\rho_{j+1/2}\longleftrightarrow 4\pi r_{i+1/2}^{2}\rho_{i+1/2}=%
\frac{m_{i+1}-m_{i}}{\Delta}  \label{md}
\end{equation}
Note that the exact linear relation between the masses and densities means
that these variables are completely equivalent. Similarly, one has that%
\begin{equation}
\frac{\partial}{\partial m_{i}}=\frac{1}{\Delta4\pi}\left( \frac{1}{%
r_{i-1/2}^{2}}\frac{\partial}{\partial\rho_{i-1/2}}-\frac{1}{r_{i+1/2}^{2}}%
\frac{\partial}{\partial\rho_{i+1/2}}\right) .  \label{m1}
\end{equation}
With this discretization, the variable $\rho_{N+1/2}$ is the concentration
at the wall and $m_{N+1}$ is the total mass within the container. Since mass
is conserved, the value of $m_{N+1}$ is, by definition, constant and that of 
$\rho_{N+1/2}$ is determined by mass conservation and the other densities
via Eq.(\ref{md}).

\subsubsection{The free energy}

The free energy is discretized using the same integration scheme as relates
the densities and masses, 
\begin{equation}
\beta F=\Delta \sum_{j=0}^{\infty }4\pi r_{j+1/2}^{2}\beta f\left( \rho
_{j+1/2}\right) +\Delta \sum_{j=0}^{\infty }4\pi \frac{1}{2}K\left( r\frac{%
\partial \rho }{\partial r}\right) _{j+1/2}^{2}
\end{equation}%
Note that the upper limits on the sums are infinite. This is because a
boundary condition on the concentration is needed in order to be able to
evaluate the square-gradient term at the boundary of the container. To this
end, it will be assumed that \textit{outside} the container, there is a
uniform concentration denoted $\rho _{\infty }$ so that, formally, $\rho
_{i+1/2}=\rho _{\infty }$ for $i\geq N$. This concentration could be zero,
in the case of a truly isolated system, or could be set to a finite value in
order, e.g., to approximate nucleation in an infinite system. As it stands,
the squared gradient term requires the concentration derivative at
intermediate points. This is approximated as 
\begin{equation}
\left( r\frac{\partial \rho }{\partial r}\right) _{j+1/2}^{2}=\frac{1}{2}%
\left( \left( r\frac{\partial \rho }{\partial r}\right) _{j}^{2}+\left( r%
\frac{\partial \rho }{\partial r}\right) _{j+1}^{2}\right) +O\left( \Delta
\right)
\end{equation}%
giving%
\begin{align}
\beta F& =\Delta \sum_{j=0}^{\infty }4\pi r_{j+1/2}^{2}\beta f\left( \rho
_{j+1/2}\right) +\Delta \sum_{j=0}^{\infty }2\pi K\left( r\frac{\partial
\rho }{\partial r}\right) _{j}^{2} \\
& =\Delta \sum_{j=0}^{\infty }4\pi r_{j+1/2}^{2}\beta f\left( \rho
_{j+1/2}\right) +\Delta \sum_{j=0}^{\infty }2\pi Kr_{j}^{2}\left( \frac{\rho
_{j+1/2}-\rho _{j-1/2}}{\Delta }\right) ^{2}  \notag
\end{align}%
or, with the boundary condition,%
\begin{equation}
\beta F-\beta F_{\infty }=\Delta \sum_{j=0}^{N}4\pi r_{j+1/2}^{2}\beta
f\left( \rho _{j+1/2}\right) +\Delta \sum_{j=0}^{N+1}2\pi K r_{j}^{2}\left( 
\frac{\rho _{j+1/2}-\rho _{j-1/2}}{\Delta }\right) ^{2}  \label{FE}
\end{equation}%
where $\beta F_{\infty }$ is the energy of the system with uniform
concentration $\rho _{\infty }$.

\subsubsection{The stochastic differential equation}

With this discretization, the SDE becomes%
\begin{equation}
\frac{\partial m_{i}\left( t\right) }{\partial t}=-D\frac{m_{i+1}\left(
t\right) -m_{i-1}\left( t\right) }{2\Delta}\frac{\partial F}{\partial\Delta
m_{i}\left( t\right) }+\sqrt{2D\frac{m_{i+1}\left( t\right) -m_{i-1}\left(
t\right) }{2\Delta}}\sqrt{\frac{1}{\Delta}}\mathbf{\xi}_{i}\left( t\right)
,\;\;1\leq i\leq N  \label{sd0}
\end{equation}
with $\left\langle \mathbf{\xi}_{i}\left( t\right) \mathbf{\xi}_{j}\left(
t^{\prime}\right) \right\rangle =\delta\left( t-t^{\prime}\right)
\delta_{ij} $. There are only equations for $m_i$ for $i$ up to $N$ since,
as discussed above, the wall is at position $N+1/2$ and the total mass in
the container, $m_{N+1}$, is conserved. Using Eq.(\ref{md}) and (\ref{m1})
gives 
\begin{equation}
\frac{d\rho_{i+1/2}}{dt}=\frac{1}{4\pi r_{i+1/2}^{2}\Delta}\left( \left(
1-\delta_{iN}\right) \frac{dm_{i+1}}{dt}-\left( 1-\delta_{i0}\right) \frac{%
dm_{i}}{dt}\right) ,\;\;0\leq i\leq N.
\end{equation}
where it the invariance of $m_{N+1}$ and of $m_{0}=0$ have been explicitly
indicated. Expanding, this gives, 
\begin{equation}  \label{sded}
\frac{\partial\rho_{i+1/2}\left( t\right) }{\partial t}=-D%
\sum_{j=0}^{N}g_{ij}^{-1}\frac{\partial F}{\partial\rho_{j+1/2}\left(
t\right) }+\sqrt{2D}\sum_{j=0}^{N}q_{ij}^{-1}\xi_{j}\left( t\right)
,\;\;0\leq i\leq N
\end{equation}
with%
\begin{align}
g_{ij}^{-1} & =\left( 1-\delta_{iN}\right) \frac{r_{i+3/2}^{2}\rho
_{i+3/2}+r_{i+1/2}^{2}\rho_{i+1/2}}{8\pi r_{i+1/2}^{2}\Delta^{3}}\left( 
\frac{1}{r_{i+1/2}^{2}}\delta_{ij}-\frac{1}{r_{i+3/2}^{2}}\delta
_{i+1j}\right) \\
& -\left( 1-\delta_{i0}\right) \frac{r_{i+1/2}^{2}\rho_{i+1/2}+r_{i-1/2}^{2}%
\rho_{i-1/2}}{8\pi r_{i+1/2}^{2}\Delta^{3}}\left( \frac{1}{r_{i-1/2}^{2}}%
\delta_{i-1j}-\frac{1}{r_{i+1/2}^{2}}\delta_{ij}\right) ,\;\;0\leq i,j\leq N
\notag
\end{align}
and%
\begin{equation}
q_{ij}^{-1}=\frac{1}{4\pi\sqrt{2}\Delta^{3/2}}\frac{\left( 1-\delta
_{iN}\right) \delta_{i+1j}-\left( 1-\delta_{i0}\right) \delta_{ij}}{%
r_{i+1/2}^{2}}\sqrt{r_{j+1/2}^{2}\rho_{j+1/2}+r_{j-1/2}^{2}\rho_{j-1/2}}%
,\;\;0\leq i,j\leq N
\end{equation}
It is easy to check that the fluctuation-dissipation relation $\mathbf{g}%
^{-1}=\mathbf{qq}^{T}$ holds and that the discretized equation remains Ito-Stratonovich equivalent\cite{Gardiner}.

Just as one of the mass variables is fixed by the constraint of constant
mass, so one of the concentration variables is also redundant. In fact, from
the previous equation it follows, since the total mass is conserved, that%
\begin{equation}
\frac{d\rho_{N+1/2}}{dt}=\frac{1}{4\pi r_{i+1/2}^{2}\Delta}\left( \frac{%
dm_{N+1}}{dt}-\frac{dm_{N}}{dt}\right) =-\frac{1}{4\pi r_{i+1/2}^{2}\Delta}%
\frac{dm_{N}}{dt}
\end{equation}
There are two possibilities:\ either one proceeds using this as the equation
of motion or one eliminates the variable $\rho_{N+1/2}$ altogether by means
of the expression for the total mass giving%
\begin{equation}
r_{N+1/2}^{2}\rho_{N+1/2}=\frac{1}{4\pi}m_{N+1}-%
\sum_{i=0}^{N-1}r_{i+1/2}^{2}\rho_{i+1/2}
\end{equation}
so that in fact the remaining variables are the masses for $1\leq j\leq N$
or, equivalently, the densities up to $\rho_{N-1/2}$. When evaluating
derivatives, one must then use, e.g., 
\begin{align}
\frac{\partial\beta F}{\partial\rho_{i+1/2}} & =\left. \frac{\partial\beta F%
}{\partial\rho_{i+1/2}}\right\vert _{\rho_{N+1/2}}+\frac{\partial\rho
_{N+1/2}}{\partial\rho_{i+1/2}}\frac{\partial\beta F}{\partial\rho_{N+1/2}}
\label{E} \\
& =\left. \frac{\partial\beta F}{\partial\rho_{i+1/2}}\right\vert
_{\rho_{N+1/2}}-\frac{r_{i+1/2}^{2}}{r_{N+1/2}^{2}}\frac{\partial\beta F}{%
\partial\rho_{N+1/2}}  \notag
\end{align}
The notation indicates that in the first term on the right hand side the
derivatives are evaluated holding $\rho_{N+1/2}$, as well as all other
concentration $\rho_{j+1/2}$ except for $j=i$, constant. Note that the
second term on the right plays no role when discretizing the SDE, Eq.(\ref%
{sded}).

\subsubsection{Stationary solutions}

A special role will be played by the stationary solutions to the
deterministic dynamics (i.e. the dynamics with no noise). These will include
both the metastable initial and final states and the metastable critical
cluster separating them. They are simply the solutions of 
\begin{equation}
\frac{\partial F}{\partial m_{i}}=0,\;\;1\leq i\leq N
\end{equation}
under the constraint that the total mass is conserved (so that all
derivatives are understood to be evaluated with $m_{N+1}$ fixed).
Equivalently, in terms of the concentration this gives%
\begin{equation}
\frac{1-\delta_{iN}}{4\pi r_{i+1/2}^{2}}\frac{\partial F}{\partial\rho
_{i+1/2}}-\frac{1}{4\pi r_{i-1/2}^{2}}\frac{\partial F}{\partial\rho_{i-1/2}}%
=0,\;1\leq i\leq N.
\end{equation}
where the derivatives are evaluated as indicated in Eq.(\ref{E}), so that
this could also be written as%
\begin{align}
\frac{1}{4\pi r_{i-1/2}^{2}}\left. \frac{\partial F}{\partial\rho_{i-1/2}}%
\right\vert _{\rho_{N+1/2}} & =\frac{1}{4\pi r_{i+1/2}^{2}}\left. \frac{%
\partial F}{\partial\rho_{i+1/2}}\right\vert _{\rho_{N+1/2}},\;1\leq i\leq
N-1 \\
\frac{1}{4\pi r_{N-1/2}^{2}}\left. \frac{\partial F}{\partial\rho_{N-1/2}}%
\right\vert _{\rho_{N+1/2}} & =\frac{1}{4\pi r_{N+1/2}^{2}}\frac{\partial F}{%
\partial\rho_{N+1/2}}.  \notag
\end{align}
Finally, by introducing an additional variable, $\mu$, these $N$ equations
can be replaced by the $N+1$ equations%
\begin{align}  \label{ec}
\left. \frac{\partial F}{\partial\rho_{i-1/2}}\right\vert _{\rho_{j+1/2}} &
=4\pi r_{i-1/2}^{2}\mu,\;1\leq i\leq N \\
m_{N+1} & =\Delta\sum_{j=0}^{N}4\pi r_{j+1/2}^{2}\rho_{j+1/2}  \notag
\end{align}
for the variables $\rho_{i+1/2}$, $0\leq i\leq N$ and $\mu$. In this form,
it is clear that the stationary solutions simply correspond to extrema of
the free energy functional subject to the condition of constant mass, as one
would expect. A convenient method of solution of these equations is given in
the Appendix \ref{App1}.

\subsection{Nucleation pathway}

In principle, one could simulate the discretized SDE but the goal here is to
more directly characterize the nucleation pathway. Since the problem of
interest is the nucleation of a dense phase, with average bulk concentration 
$\rho_{l}$, from a dilute phase with average bulk concentration $\rho_{v}$,
one approach to characterizing the nucleation pathway is to look for the
most likely path\cite{Lutsko_JCP_2011_Com, Lutsko_JCP_2012_1,
Lutsko_JCP_2012_2} connecting an initial state with $\rho\left( r;t=0\right)
\sim\rho_{v}$ and final state $\rho\left( r;t=T\right) \sim\rho_{l}$. For an
infinite system, these conditions would be strict equivalences but for a
finite system, there will in general be spatial variation of the
concentration, even in the initial state due to the boundaries. Furthermore,
for a finite container, since mass is conserved, there cannot be a complete
conversion from the initial state $\rho\left( r;t=0\right) \sim\rho_{v}$ to
the final concentration, $\rho_{l}$:\ the lowest energy state that conserves
mass will consist of a droplet inside of which the concentration is roughly
that of the final state, $\rho_{l}$, and outside of which the concentration
decays to some value that may or may not be near $\rho_{v}$ depending on the
boundary conditions.

It has previously been shown that in the weak noise limit, in which the
thermodynamic driving force is large compared to typical noise fluctuations,
the most likely path connecting the two metastable states will pass through
the critical cluster\cite{Lutsko_JCP_2011_Com,Lutsko_JCP_2012_1}.
Furthermore, the MLP can be determined by following the deterministic
equations of motion forward and backward in time. In practice, this means
that one starts at the critical cluster and then perturbs the system
slightly, first in the direction of one metastable state and then in the
direction of the other, and follows the deterministic dynamics given in the
present case by 
\begin{equation}
\frac{\partial m_{i}\left( t\right) }{\partial t}=-D\frac{m_{i+1}\left(
t\right) -m_{i-1}\left( t\right) }{2\Delta }\frac{\partial F}{\partial
\Delta m_{i}\left( t\right) },\;\;1\leq i\leq N  \label{det}
\end{equation}%
or, equivalently,%
\begin{equation}
\frac{\partial \rho _{i+1/2}\left( t\right) }{\partial t}=-D%
\sum_{j=0}^{N}g_{ij}^{-1}\frac{\partial F}{\partial \rho _{j+1/2}\left(
t\right) },\;\;0\leq i\leq N-1  \label{det1}
\end{equation}%
where, again, the derivatives on the right are evaluated using Eq.(\ref{E}).
There are several important points to be made about this procedure. First,
even though there is no noise term in Eq.( \ref{det}), this is a noise
driven process:\ Eq.(\ref{det}) is simply a convenient and exact means of
determining the most likely path which is obviously a concept rooted in the
fact that the full dynamics includes fluctuations\cite{Lutsko_JCP_2011_Com,
Lutsko_JCP_2012_1}. Second, the fact that one follows the path in both
directions away from the critical cluster, or equivalently, that one follows
the dynamics forward and backwards in time, does not in any way imply
time-reversal invariance. In fact, the underlying dynamics is dissipative
and so is definitively not time-reversal invariant. Again, the fact that
this procedure serves to determine the MLP is a mathematical result that
does not imply anything concerning the time-reversal properties of the
actual stochastic dynamics. Finally, the first part of the prescription,
namely perturbing away from the critical cluster, must be made more precise.
One must perturb away from the initial state because the deterministic
driving force in the critical cluster is zero: it is an unstable stationary
point of the dynamics. Thus, in principle one must move an infinitesimal
distance away from it to initiate the dynamics, but in practise one must of
course make a finite perturbation with the understanding that the final
results are approximate and approach the exact result only as the size of
the initial perturbation goes to zero. To construct the perturbation, one
first solves a generalized eigenvalue problem as described in Appendix %
\ref{App2} which typically results in the identification of a single,
unstable eigenvalue. The associated eigenvector, $v_{i}$, defines the
unstable direction and the initial state is simply%
\begin{equation}
\rho _{i+1/2}=\rho _{i+1/2}^{\ast }\pm \epsilon v_{i},\;\;0\leq i\leq N-1
\end{equation}
where $\epsilon$ is a parameter controlling the size of the imposed perturbation.

The next step is the numerical integration of Eq.(\ref{det1}). Because of
the Ito-Stratonovich equivalence of the (spatially-)discretized SDE there is
no constraint on the discretization scheme used for the time variable. For
example, with discrete time variables $t_i=i\tau$, the left hand side is
simply $(\rho_{i+1/2}(t_{j+1})-\rho_{i+1/2}(t_j))/\tau$ while the right hand
side can be evaluated at $t_j$, giving an Euler scheme, or at $t_j+\tau/2$
giving a more efficient implicit scheme. Further details are given in
Appendix \ref{App3}.

Finally, it is convenient to introduce a parameterization for the nucleation
pathway. As described elsewhere\cite{Lutsko_JCP_2012_1, Lutsko_JCP_2012_2},
the structure of the stochastic model induces a natural measure of distance
in concentration space between the concentration at time $t_1$ and that at
time $t_2$ given by 
\begin{equation}
d[\rho_1,\rho_2]=\int_{t_1}^{t_2}\sqrt{\int_0^{\infty}\frac{1}{4\pi r^2
\rho(r;t)}\left(\frac{\partial m(r;t)}{\partial r}\right)^2dr}dt
\end{equation}
By definition, this distance increases monotonically along any proposed
nucleation pathway and therefore serves as a natural, induced reaction
coordinate.

\section{Example:\ Nucleation of dense protein solution from dilute solution}

In order to illustrate the theory developed above, detailed calculations for
a model globular protein\cite{GuntonBook} have been performed. The effect of
the solvent was approximated, crudely, by assuming Brownian dynamics of the
(large) solute molecules which also experience an effective pair interaction
for which the ten Wolde-Frenkel interaction potential\cite{tWF-Proteins}%
\begin{equation}
v\left( r\right) =\left\{ 
\begin{array}{c}
\infty ,\;\;r\leq \sigma \\ 
\,\frac{4\,\epsilon }{\alpha ^{2}}\left( \,\left( \frac{1}{(\frac{r}{\sigma }%
)^{2}-1}\right) ^{6}-\,\alpha \,\left( \frac{1}{(\frac{r}{\sigma })^{2}-1}%
\right) ^{3}\right) ,\;\;r\geq \sigma%
\end{array}%
\right.
\end{equation}%
was used with $\alpha =50$ which is then cutoff at $r_{c}=2.5\sigma $ and
shifted so that $v\left( r_{c}\right) =0$. The temperature is fixed at $%
k_{B}T=0.375\epsilon $ and the equation of state computed using
thermodynamic perturbation theory. The transition studied was that between
the dilute phase and the dense protein phase which, in the present
simplified picture, is completely analogous to the vapor-liquid transition
for particles interacting under the given pair potential. The gradient
coefficient, $K$, is calculated from the pair potential using the
approximation given in Ref. \onlinecite{Lutsko2011a} 
\begin{equation}
\beta K\simeq -\frac{2\pi }{45}d^{5}\beta v\left( d\right) +\frac{2\pi }{15}%
\int_{d}^{\infty }\left( 2d^{2}-5r^{2}\right) \beta v\left( r\right) r^{2}dr
\end{equation}%
where $d$ is the effective hard-sphere diameter for which the
Barker-Henderson approximation was used\cite{BarkerHend}. For the
temperature used here it was found that $\beta K=1.80322\sigma ^{5}$.

\begin{figure}[ptb]
\includegraphics[angle=-90,scale=0.4]{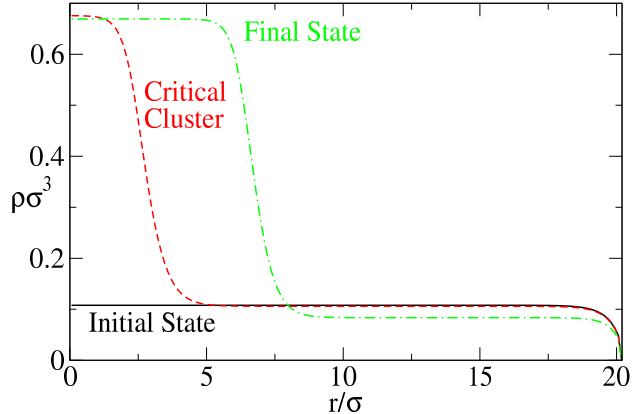} 
\caption{The concentration distribution in the metastable, nearly uniform initial state (solid line), 
the critical cluster (broken line) and the final stable cluster (dash-dot line) as a 
function of distance from the center of the cluster.} \label{fig1}
\end{figure}

The nucleation pathway has been evaluated for a spherical box with a radius
or $20.63\sigma $ using a lattice of $160$ points with a spacing equal to $%
0.126\sigma $ and with the external concentration set to zero. The time step
used was $10^{-4}D^{-1}\sigma^2$. The only remaining control parameter is
the initial mass of material in the container. For the chosen temperature,
the coexistence densities of the dilute and dense phases are $%
\rho\sigma^3=0.071$ and $\rho\sigma^3=0.663$, respectively. The initial
condition for the calculations presented here was a uniform concentration
which was $1.5$ times the dilute phase coexistence concentration which
serves to fix the total mass. The various metastable states were then
determined under the constraint that the total mass does not vary. Figure %
\ref{fig1} shows the initial metastable state, the critical cluster and the
final, stable state. Since the material in the center of the cluster that
eventually forms is near the bulk state, it is clear that the if the initial
state contains too little matter, no cluster will form. In fact, if the
amount of matter in the system is scaled with the radius, then as the radius
of the system is reduced, the final stable cluster has smaller and smaller
radius until at some point it is smaller than the critical cluster:\ i.e.,
there is no barrier separating the initial and final states.

One quantity of particular interest is the size of the cluster. In previous
work on infinite systems, this was characterized by both the equimolar
radius and by identifying the point at which the difference of the
concentration from the background was sufficiently small. In the present
case of relatively small, finite systems that conserve mass, the equimolar
radius is not very useful so only the second measure is employed. Even then,
the "background" concentration has to be reinterpreted:\ here, it is taken
to be the concentration a distance of $2\sigma $ from the wall (so as to
avoid boundary effects), a point denoted as $r_{b}$ since it defines the
"background". The radius is then determined by starting at the center of the
cluster, i.e. at $r=0$, and increasing $r$ until a point $R$ is found at
which 
\begin{equation*}
\rho \left( R\right) =\rho \left( r_{b}\right) +\epsilon \left( \rho \left(
0\right) -\rho \left( r_{b}\right) \right)
\end{equation*}%
for some prescribed value of $\epsilon $, typically either $\epsilon =\frac{1%
}{2}$ (which will be referred to as the half width) or $\epsilon =0.1$
(which will be referred to as the total width).

\begin{figure}[ptb]
\includegraphics[angle=-90,scale=0.25]{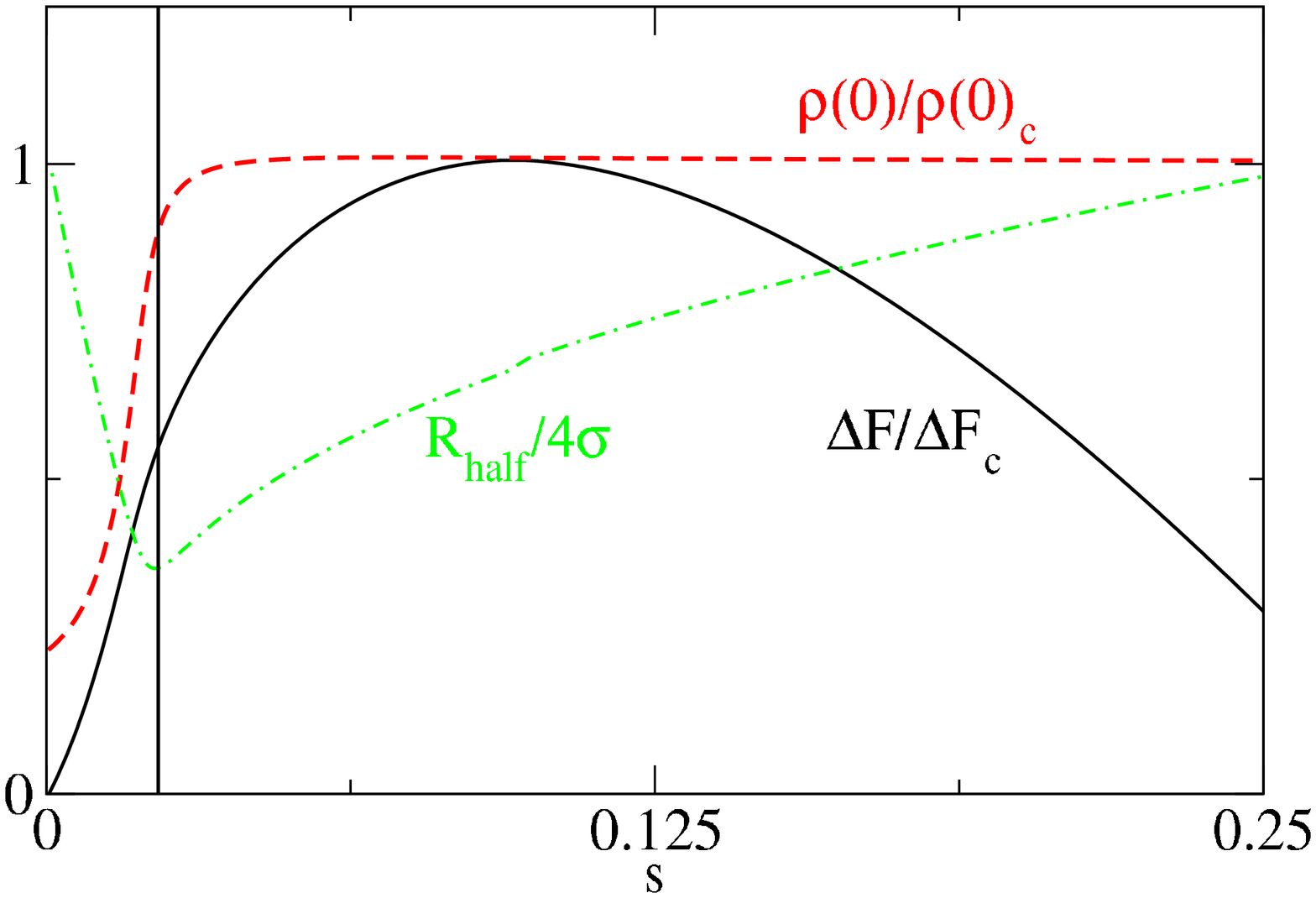} %
\includegraphics[angle=-90,scale=0.25]{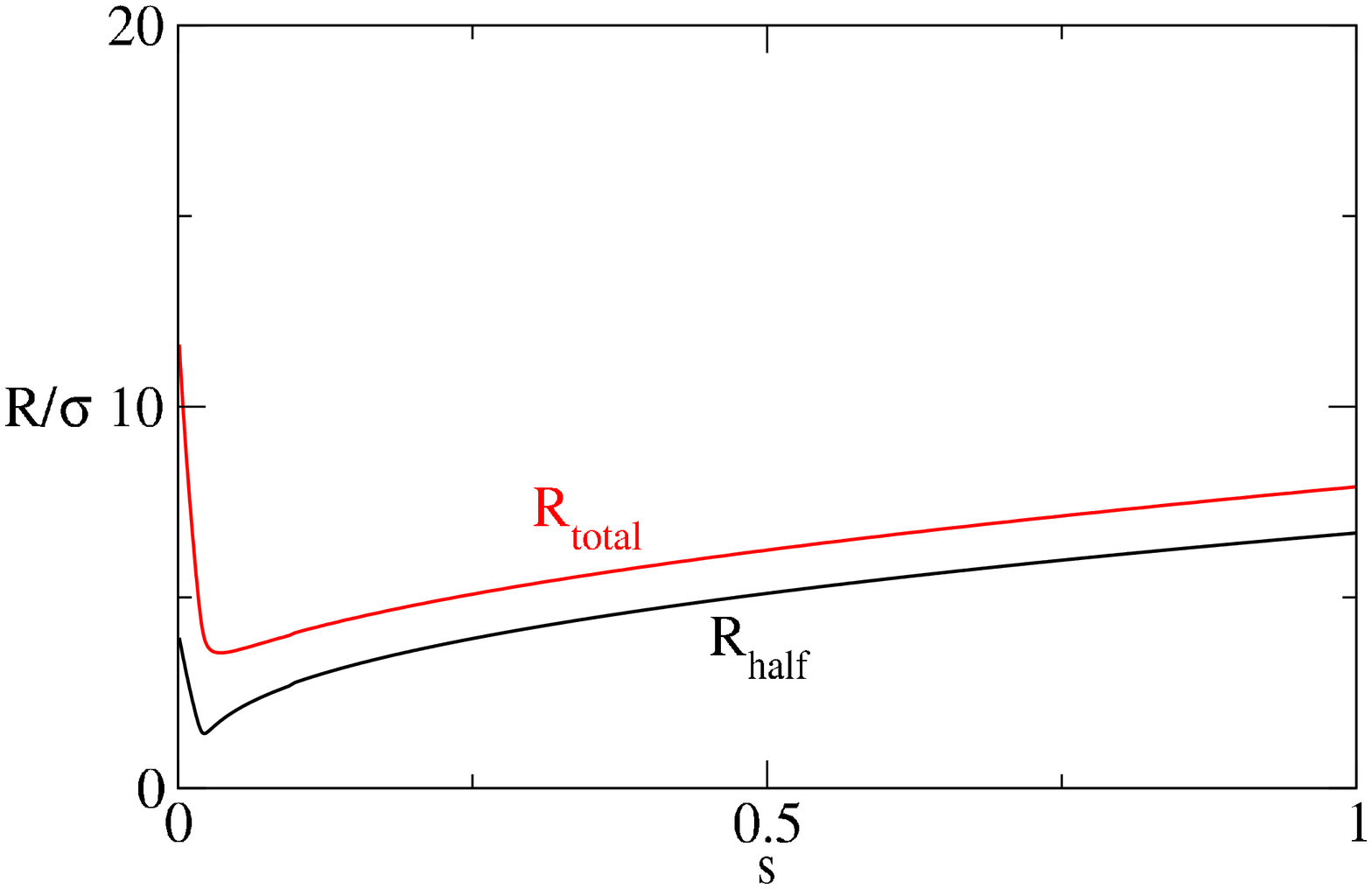} 
\caption{The left panel shows the
free energy difference from the initial state, $\Delta F/\Delta F_c$, the
concentration at the center of the cluster, $\rho(0)/\rho_c(0)$, and the
cluster radius, $R_{half}/(4\sigma)$, as functions of distance along the
nucleation pathway. The energy and concentration are scaled by their values
in the critical cluster whereas the radius has been arbitrarily scaled to
$4\sigma$. The vertical line shows the point at which the radius reaches its
minimum. The right panel shows two measures of the cluster size as a
function of distance along the pathway.} \label{fig2}
\end{figure}

The determination of the two branches of the pathway was initiated by
perturbing in the direction of the unstable eigenvector with the
proportionality constant being adjusted so as to produce a change in free
energy of $\Delta F=0.005k_{B}T$ from the critical cluster. Figure \ref{fig2}
shows the evolution of the cluster radius, the central concentration and the
free energy difference as functions of distance as well as of the cluster
size along the nucleation pathway. The results are similar to those obtained
previously for an infinite system using parameterized concentration
profiles. In particular, the fact that the process of nucleation involves
two separate steps:\ first, a long wavelength, small amplitude concentration
fluctuation forms. This stage is characterized by a large initial radius
that decreases while the central concentration increases. At a certain
moment, when the concentration is near its bulk value, the radius begins to
increase and, in this second stage, the evolution of the cluster follows the
classical scenario of monotonic growth of radius and mass at constant,
near-bulk interior concentration.

\begin{figure}[ptb]
\includegraphics[angle=-90,scale=0.25]{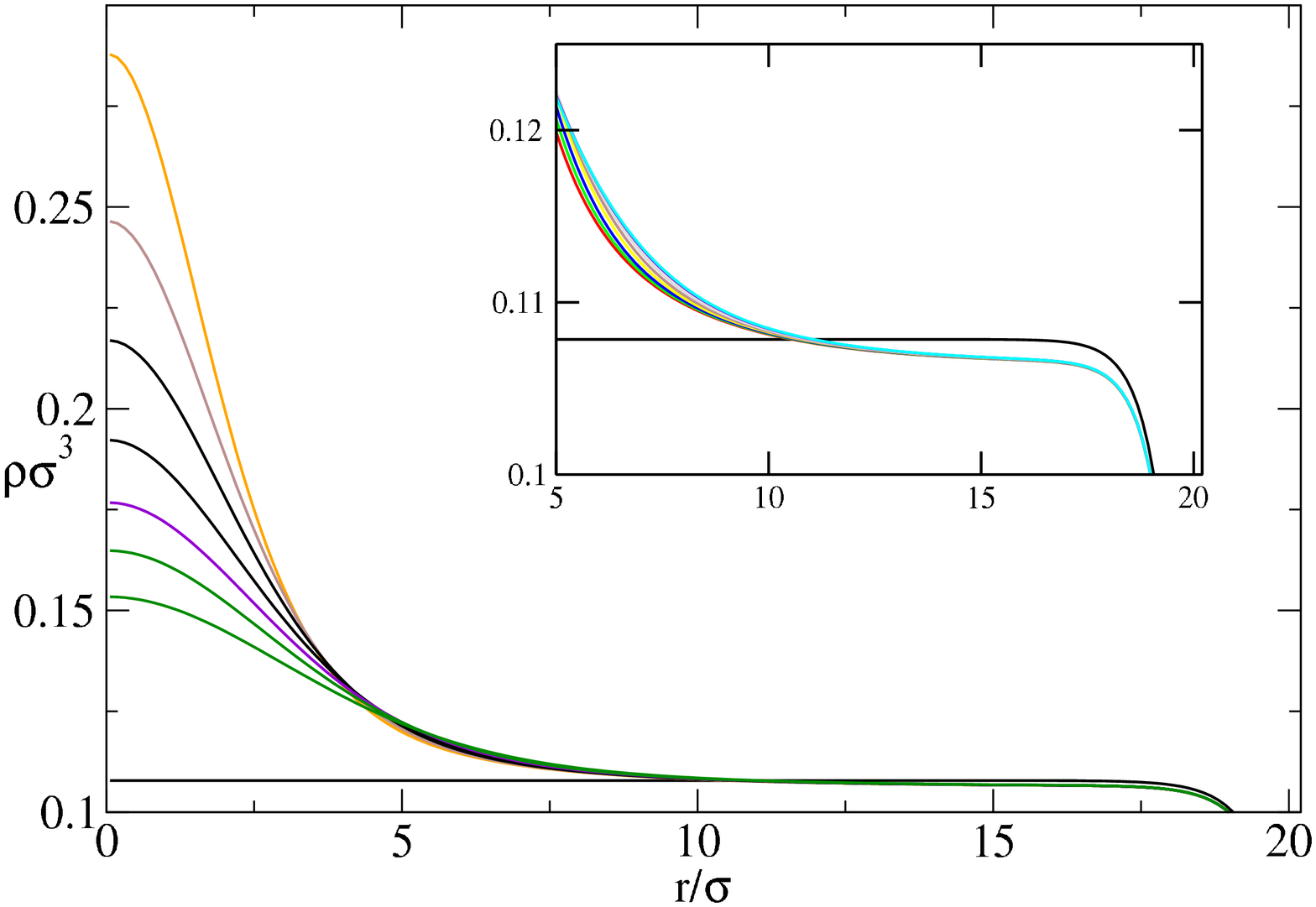} %
\includegraphics[angle=-90,scale=0.25]{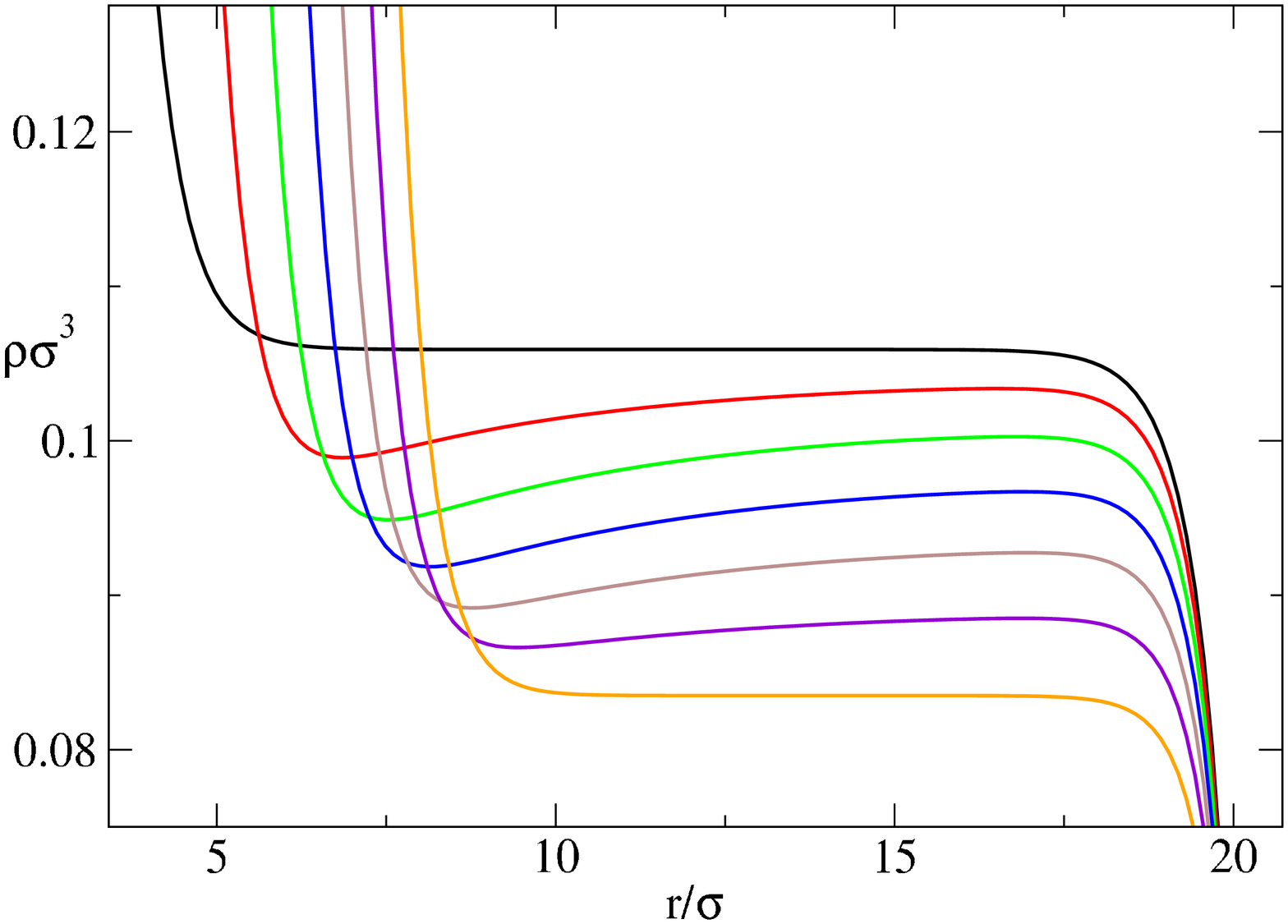} 
\caption{The evolution of
the concentration along the nucleation pathway. The left panel shows the
early stages of formation of the pre-critical cluster (increasing
concentration at $r=0$ corresponds to increasing distance along the
nucleation pathway). The large spatial extent of the initial concentration
perturbation is evident and the inset shows the monotonic variation of the
concentration as a function of distance from the center. The right panel
shows part of the concentration distribution during post-critical growth
(the maximum concentration moves to increasing radii as one progresses along
the nucleation pathway). The critical cluster, left-most curve, and final
state, right-most curve, show no depletion zone, but the presence of a
depletion zone in intermediate clusters is clear.} \label{fig3}
\end{figure}

There is one significant difference in the pre- and post-critical cluster
growth as shown in Fig. \ref{fig3}. Growth of the post-critical cluster is
deterministic and driven by the fact that the free energy of the system
decreases with increasing cluster size so that material is actively drawn
into the cluster. The fact that the low-concentration protein drawn into the
cluster can only be replaced by the diffusion of new material from further
away results in the formation of a region with a lower concentration of
protein than in the mother phase, further from the cluster; in other words,
in the formation of a depletion zone. The formation of a depletion zone is
expected in classical treatments of cluster growth\cite{Saito}. The process
is governed by the fact that the lower the concentration in the mother
phase, the larger the corresponding critical cluster. If material is drawn
into the growing cluster faster than new material can replace it by
diffusion, (a situation that is always assumed to occur for sufficiently
large clusters since the thermodynamic driving force increases with cluster
size), then the cluster will lower the concentration in the adjacent region
to the point that the thermodynamic driving force is sufficiently small that
the rate of addition of material to the cluster will be equal to the rate of
replenishment by diffusion. Assuming diffusion is slow, this will force the
cluster to be very near the critical cluster for the adjacent depleted
mother phase.

In contrast to the formation of the depletion zone for post-critical
clusters, Fig. \ref{fig3} shows that no depletion zone exists for
pre-critical clusters. The reasoning is the opposite to that for
post-critical growth:\ the pre-critical cluster is unstable and its growth
is governed by unlikely fluctuations. If the concentration outside the
cluster is unnecessarily low, then the cluster will be even more unstable
(since, as pointed out above, lower concentration in the mother phase
increases the size of the critical cluster). Hence, the most favorable way
to form an unstable cluster is one that does not create any region of
lowered concentration. Of course, since mass is conserved, the concentration
outside the cluster \textit{must} decrease as the cluster grows, but, as can
be seen in the figure, the concentration is at a maximum just "outside" the
cluster (however that is defined) and decreases monotonically as one moves
further away from it so that the concentration of the mother phase just
outside the cluster is as large as possible, i.e. the maximum of the
concentration anywhere in the mother phase. This illustrates a point made
previously\cite{Lutsko_JCP_2012_1}, namely that the initial formation of a
cluster as a localized object with small radius would necessarily involve
the formation of a depletion zone, due to mass conservation, and this would
clearly be unfavorable since it would make any (unstable) pre-critical
cluster even more unstable. Instead, what the present results show is that
the most likely nucleation pathway involves a long-wavelength concentration
fluctuation that contains and localizes enough mass to form the critical
cluster.

\section{Conclusions}

A discretized version of the stochastic field equations for fluctuations in
a fluid of Brownian particles in the over-damped limit have been given. The
most likely path for nucleation was computed in the weak-noise limit and it
was found that the pathway obeyed the same two-stage mechanism as previously
found for infinite systems. Namely, nucleation begins with a
long-wavelength, small-amplitude fluctuation that evolves by becoming
spatially more localized until the concentration inside the fluctuation
reaches near-bulk values. This is followed by a second stage of spatial
growth of the cluster that is similar to that assumed in\ CNT.

It was observed that an important difference between pre-critical and
post-critical growth is that in the former case, no depletion zone forms
whereas in the latter, a depletion zone is clearly in evidence. The
existence of depletion zone is not surprising and, indeed, is assumed in
classical treatments of cluster growth. However, its presence is important
in distinguishing the results of the present, dynamical approach to
nucleation from those of other approaches. One alternative method is the use
of equilibrium DFT. The determination of the properties of the critical
cluster via DFT is relatively uncontroversial, however the use of DFT to
determine the pathway, while widely practiced, is harder to justify. This is
because DFT is a strictly equilibrium theory\cite{LutskoAdvChemPhysDFT}
whereas nucleation is, by its nature, a nonequilibrium, fluctuation-driven
processes\cite{Lutsko_JCP_2012_1}. It is therefore unclear why DFT should be
the right tool to describe the nucleation pathway.\ This is reflected in the
fact that it is difficult to map out the pathway using DFT which is based on
minimizing a free energy functional. The problem is that non-critical
clusters are obviously not stationary points of the free energy so that one
can only determine their properties using DFT if some sort of external
constraint is applied\cite{Corti_PRL, Ghosh, EvansArcherNucleation} so as to
stabilize them. This leads to the problem that there is no unique constraint
and different constraints can lead to different results\cite{LutskoBubble1,
Lutsko2011a}.

In answer to this problem, another approach has gained popularity in recent
years. It involves the determination of the minimum free energy pathway
connecting the critical cluster to the initial and final states\cite%
{LutskoBubble1, LutskoBubble2, Philippe, MFEP_App, LutskoAdvChemPhysCrys,
MFEP_App, Nucleation_Copolymer}. While this does not involve the imposition
of any constraints, it nevertheless suffers from a similar arbitrariness as
the MFEP\ can only be calculated once a method for determining distances in
concentration-space has been adopted\cite{LutskoBubble2}. Again, without
further input the choice of metric in concentration space is arbitrary and
the method ends up being no better founded than the older constraint-based
methods.

Aside from these criticisms of the internal logic of the DFT\ methods, the
present results show a further qualitative failing that demonstrates that
free energy considerations, without dynamics, cannot be enough. As discussed
above, the formation of a depletion zone is an inevitably consequence of the
fact that the thermodynamic driving force for the incorporation of new
material into the cluster grows as the cluster grows and must, at some
point, cause material to be drawn into the cluster faster than it can be
diffusively replaced. None of the DFT studies report seeing a depletion zone
for super critical clusters. In all cases, the concentration seems to decay
monotonically to the surrounding bulk. The is a clear sign that approaches
that do not include dynamics are not sufficient.

An alternative to DFT is its dynamical cousin, Dynamical DFT\ (DDFT)\cite%
{Evans79, LutskoAdvChemPhysDFT}. The equations for DDFT are in fact the same
as those given above, Eq.(\ref{D0}), if the noise term is dropped or Eq.(\ref%
{D1}) if it is understood that only forward evolution in time is relevant
(for DDFT). The latter point is important:\ DDFT can only describe a system
falling down a free energy surface: by its nature, since it involves no
fluctuations, DDFT cannot describe barrier crossing. To overcome this, some
studies\cite{Archer} have relied on the inclusion of fluctuations in the
initial condition for a DDFT calculation but at best this describes a
physical system perturbed by an initial, external shock. The results will
necessarily be dependent on, e.g., the amplitude and spectrum of the noise
and thus is subject to a similar arbitrarity to that of static DFT and for
the same reason:\ both theories are based on equilibrium physics and so
cannot naturally describe nonequilibrium states. On the other hand, DDFT
does have the virtue that the post-critical growth is correctly described
and it will certainly correctly describe the depletion zone.

The dynamical theory presented here and elsewhere does, in some sense,
provide an extension of DDFT to barrier crossing problems since the MLP\ is
described by DDFT-like equations. However, there are still conceptual
differences that separate the two approaches. The most important is the
nature of the "free energy" that enters into the dynamical equations. In
DDFT, the free energy typically enters via a local equilibrium
approximation. Because of the context (i.e. that one is dealing with
ensemble-averaged quantities) this is the true, macroscopic free energy
functional. In the dynamical approach, the status of the "free energy" is
more problematic. It is certainly not the macroscopic free energy. In fact,
let us write the DDFT equations in the form%
\begin{equation*}
\frac{\partial \left\langle \rho \left( \mathbf{r};t\right) \right\rangle }{%
\partial t}=D\mathbf{\nabla }\left\langle \rho \left( \mathbf{r};t\right)
\right\rangle \mathbf{\nabla }\left( \frac{\delta \widetilde{F}\left[ \rho %
\right] }{\delta \rho \left( \mathbf{r}\right) }\right) _{\rho \left( 
\mathbf{r}\right) \rightarrow \left\langle \rho \left( \mathbf{r};t\right)
\right\rangle }
\end{equation*}%
where the brackets $\left\langle ...\right\rangle $ indicate an ensemble
average and where $\widetilde{F}\left[ \rho \right] $ is the equilibrium DFT
functional. On the other hand, the noise-averaged stochastic equations, Eq.(%
\ref{D0}), are%
\begin{equation*}
\frac{\partial \left\langle \left\langle \rho \left( \mathbf{r};t\right)
\right\rangle \right\rangle }{\partial t}=D\mathbf{\nabla }\left\langle
\left\langle \rho \left( \mathbf{r};t\right) \mathbf{\nabla }\left( \frac{%
\delta F\left[ \rho \right] }{\delta \rho \left( \mathbf{r}\right) }\right)
_{\rho \left( \mathbf{r}\right) \rightarrow \rho \left( \mathbf{r};t\right)
}\right\rangle \right\rangle
\end{equation*}%
where the double brackets $\left\langle \left\langle ...\right\rangle
\right\rangle $ indicate an average over the noise. Technically, one cannot
identify the ensemble-averaged concentration, $\left\langle \rho \left( 
\mathbf{r};t\right) \right\rangle $, with the noise-averaged concentration $%
\left\langle \left\langle \rho \left( \mathbf{r};t\right) \right\rangle
\right\rangle $ because the noise-averaged concentration also involves a
coarse-graining. However, if the spatial variations are sufficiently weak,
then it is reasonable to expect these quantities will be more or less the
same and so one expects the right hand sides of these equations to become
the same. It is then seen that equality of the right hand sides depends on a
series of factorization approximations,%
\begin{eqnarray*}
\left\langle \left\langle \rho \left( \mathbf{r};t\right) \mathbf{\nabla }%
\left( \frac{\delta F\left[ \rho \right] }{\delta \rho \left( \mathbf{r}%
\right) }\right) _{\rho \left( \mathbf{r}\right) \rightarrow \rho \left( 
\mathbf{r};t\right) }\right\rangle \right\rangle &\sim &\left\langle
\left\langle \rho \left( \mathbf{r};t\right) \right\rangle \right\rangle
\left\langle \left\langle \mathbf{\nabla }\left( \frac{\delta F\left[ \rho %
\right] }{\delta \rho \left( \mathbf{r}\right) }\right) _{\rho \left( 
\mathbf{r}\right) \rightarrow \rho \left( \mathbf{r};t\right) }\right\rangle
\right\rangle \\
&\sim &\left\langle \left\langle \rho \left( \mathbf{r};t\right)
\right\rangle \right\rangle \mathbf{\nabla }\left( \frac{\delta \widetilde{F}%
\left[ \rho \right] }{\delta \rho \left( \mathbf{r}\right) }\right) _{\rho
\left( \mathbf{r}\right) \rightarrow \left\langle \left\langle \rho \left( 
\mathbf{r};t\right) \right\rangle \right\rangle }
\end{eqnarray*}%
Interestingly, the assumption of weak noise, which underlies the development
of the dynamical approach, might be expected to justify this series of
approximations so that a certain consistency between the dynamical theory
and DDFT might be seen to emerge. In any case, it is worth noting that in
the theory of critical phenomena, it is generally accepted that a mean field
approximation of the form hard-core plus mean-field treatment of the
attractive part of the potential is a reasonable approximation for the
coarse-grained free energy\cite{Zinn-Justin}. Since this is at present the
dominant model used in DFT calculations\cite{LutskoAdvChemPhysDFT}, it might
be argued that the use of this model for the coarse-grained free energy is
actually more justifiable than is its use in DFT\ calculations.

\begin{acknowledgments}
This work was partially supported in part by the European Space Agency under
contract number ESA AO-2004-070 and by FNRS Belgium under contract C-Net
NR/FVH 972.
\end{acknowledgments}

\appendix{}

\section{Solution for the critical cluster\label{App1}}

With the discretized free energy, Eq.(\ref{FE}), the equations for the
critical cluster, Eq.(\ref{ec}), become%
\begin{align}
r_{1/2}^{2}\beta f^{\prime }\left( \rho _{1/2}\right) +Kr_{1}^{2}\left( 
\frac{\rho _{3/2}-\rho _{1/2}}{\Delta }\right) & =r_{1/2}^{2}\mu  \\
r_{i+1/2}^{2}\beta f^{\prime }\left( \rho _{i+1/2}\right) +Kr_{i}^{2}\left( 
\frac{\rho _{i+1/2}-\rho _{i-1/2}}{\Delta }\right) -Kr_{i+1}^{2}\left( \frac{%
\rho _{i+3/2}-\rho _{i+1/2}}{\Delta }\right) & =r_{1/2}^{2}\mu ,\;1\leq
i\leq N-1  \notag \\
r_{N+1/2}^{2}\beta f^{\prime }\left( \rho _{N+1/2}\right) +Kr_{N}^{2}\left( 
\frac{\rho _{N+1/2}-\rho _{N-1/2}}{\Delta }\right) -Kr_{N+1}^{2}\left( \frac{%
\rho _{\infty }-\rho _{N+1/2}}{\Delta }\right) & =r_{N+1/2}^{2}\mu   \notag
\end{align}%
and%
\begin{equation}
m_{N+1}=4\pi \sum_{i=0}^{N}r_{i+1/2}^{2}\rho _{i+1/2}.
\end{equation}%
To solve these, one begins with a guess, say $\rho _{i+1/2}$, and denote the
actual solution by $\rho _{i+1/2}^{\ast }$. Defining 
\begin{align}
x_{i}& =\rho _{i+1/2}^{\ast }-\rho _{i+1/2},\;0\leq i\leq N \\
x_{N+1}& =\mu ^{\ast }-\mu   \notag
\end{align}%
and%
\begin{align}
y_{i}& =\beta f^{\prime }\left( \rho _{i+1/2}\right) +\frac{K}{%
r_{i+1/2}^{2}\Delta }\left( r_{i}^{2}\left( \rho _{i+1/2}-\rho
_{i-1/2}\right) -r_{i+1}^{2}\left( \rho _{i+3/2}-\rho _{i+1/2}\right)
\right) ,\;0\leq i\leq N-1 \\
y_{N}& =\beta f^{\prime }\left( \rho _{N+1/2}\right) +\frac{K}{r_{N+1/2}^{2}}%
\left( r_{N}^{2}\left( \rho _{N+1/2}-\rho _{N-1/2}\right) -r_{N+1}^{2}\left(
\rho _{\infty }-\rho _{N+1/2}\right) \right)   \notag \\
y_{N+1}& =4\pi \sum_{i=0}^{N}r_{i+1/2}^{2}\rho _{i+1/2}-m_{N+1}  \notag
\end{align}%
and expanding to first order gives%
\begin{multline} 
-\frac{K}{\Delta }\left( \frac{r_{i}}{r_{i+1/2}}\right) ^{2}x_{i-1}+\left(
\beta f^{\prime \prime }\left( \rho _{i+1/2}\right) +\frac{K}{%
r_{i+1/2}^{2}\Delta }\left( r_{i}^{2}+r_{i+1}^{2}\right) \right) x_{i} \\
-\left( 1-\delta _{iN}\right) \frac{K}{\Delta }\left( \frac{r_{i+1}}{%
r_{i+1/2}}\right) ^{2}x_{i+1}-y_{N+1}=\mu -y_{i},\;0\leq i\leq N  \label{it0}
\end{multline}%
and%
\begin{equation}
4\pi \sum_{i=0}^{N}r_{i+1/2}^{2}x_{i+1/2}=y_{N+1}.  \label{it1}
\end{equation}%
These equations are tridiagonal except for the last row and column and can
be efficiently solved by a trivial modification of the Thomas algorithm\
(see, e.g., Ref. \cite{NR}). Thus, the solution to these equations is found
by starting with the boundary conditions, $\rho _{\infty }$ and $m_{N+1}$,
and an initial guess, $\rho _{i+1/2}$ and $\mu $ (for the latter, a
reasonable guess is $f^{\prime }\left( \rho _{\infty }\right) $) and then
iterating this linear system to stability. Alternatively, one can simply
choose a value for $\mu $, eliminate $y_{N+1}$ as a variable (i.e. set $%
y_{N+1}=0$ in Eq.(\ref{it0})) and drop  Eq.(\ref{it1}) and solve for the
values of $x_{i}$. The value of the mass is then evaluated post hoc.

\section{The unstable direction\label{App2}}

To determine the unstable direction, we first note that the deterministic
dynamics given in Eq.(\ref{det1}) can be interpreted as being
gradient-driven motion in a non-Euclidean geometry\cite{LutskoBubble1,
LutskoBubble2, Wales}. As such, the unstable direction is determined from
the generalized eigenvalue problem 
\begin{equation}
\sum_{j=0}^{N-1}\frac{\partial^{2}F}{\partial\rho_{i+1/2}\partial\rho_{j+1/2}%
}=\lambda\sum_{j=0}^{N-1}g_{ij}v_{j},\;\;0\leq i\leq N-1
\end{equation}
and for convenience we write%
\begin{equation}
g_{ij}^{-1}=-\frac{1}{8\pi\Delta^{3}}R_{ii^{\prime}}^{-1}\overline {g}%
_{i^{\prime}j^{\prime}}^{-1}R_{j^{\prime}j}^{-1}
\end{equation}
with%
\begin{equation}
R_{ij}\equiv\delta_{ij}r_{i+1/2}^{2}
\end{equation}
and%
\begin{align}
\overline{g}_{ij}^{-1} & =\left( 1-\delta_{iN}\right) \left(
r_{i+3/2}^{2}\rho_{i+3/2}+r_{i+1/2}^{2}\rho_{i+1/2}\right) \left(
\delta_{ij}-\delta_{i+1j}\right) \\
& -\left( 1-\delta_{i0}\right) \left(
r_{i+1/2}^{2}\rho_{i+1/2}+r_{i-1/2}^{2}\rho_{i-1/2}\right) \left(
\delta_{i-1j}-\delta_{ij}\right) ,\;\;0\leq i,j\leq N-1  \notag
\end{align}
which we also write as%
\begin{equation}
\overline{g}_{ij}^{-1}=-\left( a_{i}\delta_{ji-1}+a_{j}\delta_{ji+1}\right)
+\left( \left( 1-\delta_{i0}\right) a_{i}+a_{i+1}\right) \delta
_{ij},\;\;0\leq i,j\leq N-1
\end{equation}
where 
\begin{equation}
a_{i}=r_{i+1/2}^{2}\rho_{i+1/2}+r_{i-1/2}^{2}\rho_{i-1/2},\;\;0\leq i,j<N-1
\end{equation}
The inverse of $\overline{\mathbf{g}}^{-1}$ is easily found to be 
\begin{equation}
\overline{g}_{ij}=\sum_{l=\max(i,j)}^{N-1}\frac{1}{a_{l+1}}.
\end{equation}
Thus completing specification of the generalized eigenvalue problem.

\section{Integration scheme}

\label{App3} The deterministic equations can be written as 
\begin{equation}
\frac{d\rho_{i+1/2}}{dt}=y_{i+1/2}\left(
\rho_{i-3/2},\rho_{i-1/2},\rho_{i+1/2},\rho_{i+3/2},\rho_{i+5/2}\right)
\end{equation}
with%
\begin{align}
& y_{i+1/2}\left(
\rho_{i-3/2},\rho_{i-1/2},\rho_{i+1/2},\rho_{i+3/2},\rho_{i+5/2}\right) \\
& =-\left( 1-\delta_{N}\right) \frac{r_{i+3/2}^{2}\rho_{i+3/2}+r_{i+1/2}^{2}%
\rho_{i+1/2}}{8\pi r_{i+1/2}^{2}\Delta^{3}}\left( \frac{1}{r_{i+1/2}^{2}}%
\frac{\partial\beta F}{\partial\rho_{i+1/2}}-\frac{1}{r_{i+3/2}^{2}}\frac{%
\partial\beta F}{\partial\rho_{i+3/2}}\right)  \notag \\
& +\left( 1-\delta_{i0}\right) \frac{r_{i+1/2}^{2}\rho_{i+1/2}+r_{i-1/2}^{2}%
\rho_{i-1/2}}{8\pi r_{i+1/2}^{2}\Delta^{3}}\left( \frac{1}{r_{i-1/2}^{2}}%
\frac{\partial\beta F}{\partial\rho_{i-1/2}}-\frac{1}{r_{i+1/2}^{2}}\frac{%
\partial\beta F}{\partial\rho_{i+1/2}}\right)  \notag
\end{align}
and%
\begin{equation}
\frac{1}{r_{i+1/2}^{2}}\frac{\partial\beta F}{\partial\rho_{i+1/2}}=\Delta
4\pi\left( \beta f^{\prime}\left( \rho_{i+1/2}\right) +\frac{K}{%
r_{i+1/2}^{2}\Delta^{2}}\left[ -r_{i}^{2}\rho_{i-1/2}+\left(
r_{i}^{2}+r_{i+1}^{2}\right) \rho_{i+1/2}-r_{i+1}^{2}\rho_{i+3/2}\right]
\right)
\end{equation}
Discretizing in time with time step $\delta_{t}$ gives a simple Euler scheme,%
\begin{equation}
\rho_{i+1/2}^{\left( n+1\right) }-\rho_{i+1/2}^{\left( n\right)
}=\delta_{t}y_{i+1/2}^{\left( n\right) }
\end{equation}
with%
\begin{equation}
y_{i+1/2}^{\left( n\right) }=y_{i+1/2}\left( \rho_{i-3/2}^{\left( n\right)
},\rho_{i-1/2}^{\left( n\right) },\rho_{i+1/2}^{\left( n\right)
},\rho_{i+3/2}^{\left( n\right) },\rho_{i+5/2}^{\left( n\right) }\right)
\end{equation}
Unfortunately, this simple scheme requires a very small time step and is
therefore inefficient. As is often the case\cite{NR}, the integration can be
made much more efficient by introducing an implicit scheme. Define 
\begin{equation*}
x_{i+1/2}^{\left( n+1\right) }=\rho_{i+1/2}^{\left( n+1\right)
}-\rho_{i+1/2}^{\left( n\right) }
\end{equation*}
and evaluate the source term at the temporal "midpoint" using%
\begin{equation}
\rho_{i+1/2}^{\left( n+1/2\right) }\equiv\frac{1}{2}\left( \rho
_{i+1/2}^{\left( n+1\right) }+\rho_{i+1/2}^{\left( n\right) }\right)
=\rho_{i+1/2}^{\left( n+1\right) }-\frac{1}{2}x_{i+1/2}^{\left( n+1\right) }
\end{equation}
and expanding to first order in $x^{\left( n+1\right) }$ to get%
\begin{equation}
y_{i+1/2}^{\left( n+1/2\right) }\simeq y_{i+1/2}^{\left( n\right)
}+\sum_{j=-2}^{2}T_{ij}^{\left( n\right) }x_{j+1/2}^{\left( n+1\right) }
\end{equation}
with $T_{ij}=T_{ji}$ and 
\begin{align}
T_{ii}^{\left( n\right) } & =-\left( 1-\delta_{N}\right) \frac{1}{%
8\pi\Delta^{3}}\left( \frac{1}{r_{i+1/2}^{2}}\frac{\partial\beta F}{%
\partial\rho_{i+1/2}}-\frac{1}{r_{i+3/2}^{2}}\frac{\partial\beta F}{%
\partial\rho_{i+3/2}}\right) \\
& +\left( 1-\delta_{i0}\right) \frac{1}{8\pi\Delta^{3}}\left( \frac {1}{%
r_{i-1/2}^{2}}\frac{\partial\beta F}{\partial\rho_{i-1/2}}-\frac {1}{%
r_{i+1/2}^{2}}\frac{\partial\beta F}{\partial\rho_{i+1/2}}\right)  \notag \\
& -\left( 1-\delta_{N}\right) \frac{r_{i+3/2}^{2}\rho_{i+3/2}^{\left(
n\right) }+r_{i+1/2}^{2}\rho_{i+1/2}^{\left( n\right) }}{2r_{i+1/2}^{2}}%
\left( \frac{1}{\Delta^{2}}\beta f^{\prime\prime}\left( \rho_{i+1/2}^{\left(
n\right) }\right) +\frac{K}{\Delta^{4}}\left( \frac{r_{i}^{2}+r_{i+1}^{2}}{%
r_{i+1/2}^{2}}+\frac{r_{i+1}^{2}}{r_{i+3/2}^{2}}\right) \right)  \notag \\
& -\left( 1-\delta_{i0}\right) \frac{r_{i+1/2}^{2}\rho_{i+1/2}^{\left(
n\right) }+r_{i-1/2}^{2}\rho_{i-1/2}^{\left( n\right) }}{2r_{i+1/2}^{2}}%
\left( \frac{1}{\Delta^{2}}\beta f^{\prime\prime}\left( \rho_{i+1/2}^{\left(
n\right) }\right) +\frac{K}{\Delta^{4}}\left( \frac{r_{i}^{2}+r_{i+1}^{2}}{%
r_{i+1/2}^{2}}+\frac{r_{i}^{2}}{r_{i-1/2}^{2}}\right) \right)  \notag
\end{align}
and%
\begin{align}
T_{ii+1}^{\left( n\right) } & =-\frac{\left( 1-\delta_{N}\right)
r_{i+3/2}^{2}}{8\pi r_{i+1/2}^{2}\Delta^{3}}\left( \frac{1}{r_{i+1/2}^{2}}%
\frac{\partial\beta F^{\left( n\right) }}{\partial\rho_{i+1/2}}-\frac {1}{%
r_{i+3/2}^{2}}\frac{\partial\beta F^{\left( n\right) }}{\partial \rho_{i+3/2}%
}\right) \\
& +\left( 1-\delta_{N}\right) \frac{r_{i+3/2}^{2}\rho_{i+3/2}^{\left(
n\right) }+r_{i+1/2}^{2}\rho_{i+1/2}^{\left( n\right) }}{2r_{i+1/2}^{2}}%
\left( \frac{1}{\Delta^{2}}\beta f^{\prime\prime}\left( \rho_{i+3/2}^{\left(
n\right) }\right) +\frac{K}{\Delta^{4}}\left( \frac{r_{i+1}^{2}}{%
r_{i+1/2}^{2}}+\frac{r_{i+1}^{2}+r_{i+2}^{2}}{r_{i+3/2}^{2}}\right) \right) 
\notag \\
& +\left( 1-\delta_{i0}\right) \frac{r_{i+1/2}^{2}\rho_{i+1/2}^{\left(
n\right) }+r_{i-1/2}^{2}\rho_{i-1/2}^{\left( n\right) }}{2r_{i+1/2}^{2}}%
\frac{K}{\Delta^{4}}\left( \frac{r_{i+1}^{2}}{r_{i+1/2}^{2}}\right)  \notag
\end{align}
and%
\begin{equation}
T_{ii+2}^{\left( n\right) }=-\left( 1-\delta_{N}\right) \frac {%
r_{i+3/2}^{2}\rho_{i+3/2}^{\left( n\right) }+r_{i+1/2}^{2}\rho
_{i+1/2}^{\left( n\right) }}{r_{i+1/2}^{2}}\frac{K}{2\Delta^{4}}\left( \frac{%
r_{i+2}^{2}}{r_{i+3/2}^{2}}\right) .
\end{equation}
The integration scheme then becomes%
\begin{equation}
x_{i+1/2}^{\left( n+1\right) }=\delta_{t}y_{i+1/2}^{\left( n\right)
}+\delta_{t}\sum_{j=-2}^{2}T_{ij}^{\left( n\right) }x_{j+1/2}^{\left(
n+1\right) }
\end{equation}
Because of the banded nature of these equations, they can be solved quite
efficiently and the resulting integration scheme is much more stable than
the simple Euler scheme.

\bigskip

\bigskip

%

\end{document}